\documentclass[12pt]{article}
\usepackage{amsmath, amsthm, amssymb}
\usepackage{epsfig}
\begin{document}
\baselineskip= 24pt
\begin{center}
{\Large  Polymer Detachment Kinetics from Adsorbing Surface: Theory, Simulation
and Similarity to Infiltration into Porous Medium }\\
\vskip 0.2 true cm
\footnote{To whom correspondence should be addressed, e-mail:
jpaturej@univ.szczecin.pl}{Jaroslaw Paturej$^{1,2}$, Andrey Milchev$^{1,3}$,
Vakhtang G. Rostiashvili$^1$, and Thomas A. Vilgis$^1$} \\
\vskip 0.2true cm
{$^1$ Max Planck Institute for Polymer Research
10 Ackermannweg, 55128 Mainz, Germany\\
$^2$ Institute of Physics, University of Szczecin, Wielkopolska 15,
70451 Szczecin, Poland}\\
$^3$ Institute for Physical Chemistry, Bulgarian Academy of Science,
1113 Sofia, Bulgaria
\end{center}
\date{\today}

\begin{abstract}
The force-assisted desorption kinetics of a macromolecule from adhesive surface
is studied theoretically, using the notion of  tensile (Pincus) blobs, as well
as by means of  Monte-Carlo (MC) and Molecular Dynamics (MD) simulations. We
show that the change of detached  monomers  with time is governed by a
differential equation which is equivalent to the nonlinear porous medium
equation
(PME),  employed widely in  transport modeling of hydrogeological systems.
Depending on the pulling force and the strength of adsorption,  three kinetic
regimes can be distinguished: (i) ``trumpet'' (weak adsorption and small pulling
force), (ii) ``stem-trumpet'' (weak adsorption and moderate force), and (iii)
``stem'' (strong adsorption and large force). Interestingly, in all regimes the
number of desorbed beads $M(t)$,  and the height of the first monomer (which
experiences a pulling force) $R(t)$ above the surface follow an universal
square-root-of-time law. Consequently, the total time of detachment
$\langle \tau_d \rangle$, scales with polymer length $N$ as $\langle \tau_d
\rangle \propto N^2$.  Our main theoretical conclusions
are tested and found in agreement with data from extensive MC- and
MD-simulations.
\end{abstract}

\section{Introduction}\label{sec_intro}

During the last decade  the progress in single-molecule manipulation techniques
such as atomic force spectroscopy (AFM) and optical or magnetic tweezers
\cite{Ritort} has attracted the interest of researchers to a new field of
fascinating phenomena like DNA unzipping \cite{Somendra,Marenduzzo1,Orlandini},
or forced-induced detachment of individual polymers from adsorbing surfaces
\cite{Kierfeld}. One can experimentally test elastic and adhesive properties as
well as bond scission in polymer fibers, or in fundamental biological objects
like proteins, nucleic acids, or molecular motors with spatial resolution in the
{\em nm} range and force resolution in the {\em pN} range. In this way
substantial progress has been achieved in the understanding and use of fibers,
adhesives or biopolymers. Besides technological relevance, the field of
single-chain manipulation poses also numerous questions of interest to
fundamental physics. Thus, it has been shown, for example, that the
force-induced unzipping of the double-stranded DNA is a $1$-st order phase
transition \cite{Somendra,Marenduzzo1,Sebastian,Marenduzzo} that has the same
nature as the force-induced desorption transition of a polymer from an adhesive
substrate \cite{SBVRAMTV1,SBVRAMTV2,SBAMVRTV,Skvortsov} and that both phenomena
can be described by the same kind of {\em re-entrant} phase diagram.

While the theoretical understanding of these phase transformations in terms of
equilibrium statistical thermodynamics has meanwhile significantly improved
\cite{Skvortsov}, their kinetics has been so far less well explored. Originally,
a theoretical treatment of the problem was suggested by
Sebastian~\cite{Sebastian} while the first computer simulations of the related
phenomenon of DNA force-induced unzipping were reported by Marenduzzo {\em et
al.}~\cite{Marenduzzo}. Yet a number of basic questions pertaining to the
dependence of the total mean detachment time $\langle \tau_d \rangle$ on polymer
length $N$, pulling force $f$, or adhesion strength, $\epsilon_s$, remain open
or need validation. So far we are not aware of computer simulational studies
which would shed light on these problems. Therefore, in order to fill the gap
regarding force-induced desorption kinetics, we consider in this work a generic
problem: the force-induced detachment dynamics of a single chain, placed on
adhesive structureless plane.

The theoretical method which we used is different and more general then in Ref.
\cite{Sebastian} where a phantom (Gaussian) chain was treated in terms of the
corresponding Rouse equation. Instead, our consideration: (i) takes into account
all excluded-volume interactions among monomers and, (ii) is based on the
tensile-blob picture, proposed many years ago by Brochard-Wyart \cite{Brochard}
as theoretical framework for the description of a steady-state driven
macromolecule in solutions. This approach was generalized recently by Sakaue
\cite{Sakaue_1,Sakaue_2} for the analysis of non-equilibrium transient regimes
in polymer translocations through a narrow pore.

Very recently, Sebastian {\em et al.}~\cite{Sebastian_1} demonstrated (within
the tensile blob picture) that the average position $x(n, t)$ of a monomer $n$
at time $t$ is governed by the so-called $p$-Laplacian nonlinear diffusion
equation \cite{Simsen}. In contrast, in the present paper we discuss
force-induced desorption kinetics on the basis of a nonlinear diffusion equation
which governs the time-dependent density distribution $\rho (x, t)$ of monomers.
We show that this equation is similar to the {\it porous medium equation} (PME)
\cite{Vazquez} which has gained widespread acceptance in the theory of
groundwater transport through geological strata \cite{Barenblatt}. The most
natural way of solving the PME is based on {\it self-similarity} which implies
that the governing equation is invariant under the proper space and time
transformation \cite{Vazquez,Barenblatt}. Thus, self-similarity reduces the
problem to a nonlinear ordinary differential equation which can be solved much
more easily.

Eventually, in order to check our predictions we performed extensive MD- and
MC-simulations which, for the case of strong adsorption, large pulling
forces and overdamped regime, show an excellent agreement with our theoretical
predictions. Moreover, the comparison of MC and MD data reveals that the
presence of inertial effects, typical in underdamped dynamics, significantly
affects the observed detachment kinetics.

The paper is organized as follows: after this  short introduction, in
Section~\ref{sec_theory} we present briefly the theoretical treatment of the
problem and emphasize the different regimes pertinent to polymer detachment
kinetics. A short description of the simulational MC- and MD models is given in
Section~\ref{sec_Model} while in Section~\ref{sec_simul} we discuss the main
results provided by our computer experiments. In Section~\ref{sec_exper} we
suggest a plausible experiment which could validate our theoretical results. We
end this work by a brief summary of our findings in Section~\ref{sec_summary}.

\section{Tensile-blob picture for  the force-induced detachment}
\label{sec_theory}
\subsection{Trumpet regime}
\label{subsec_trumpet}
\subsubsection{Beads density temporal variation is described by a nonlinear
porous medium equation}
\label{sss_density}

\begin{figure}[ht]
\includegraphics[scale=0.35]{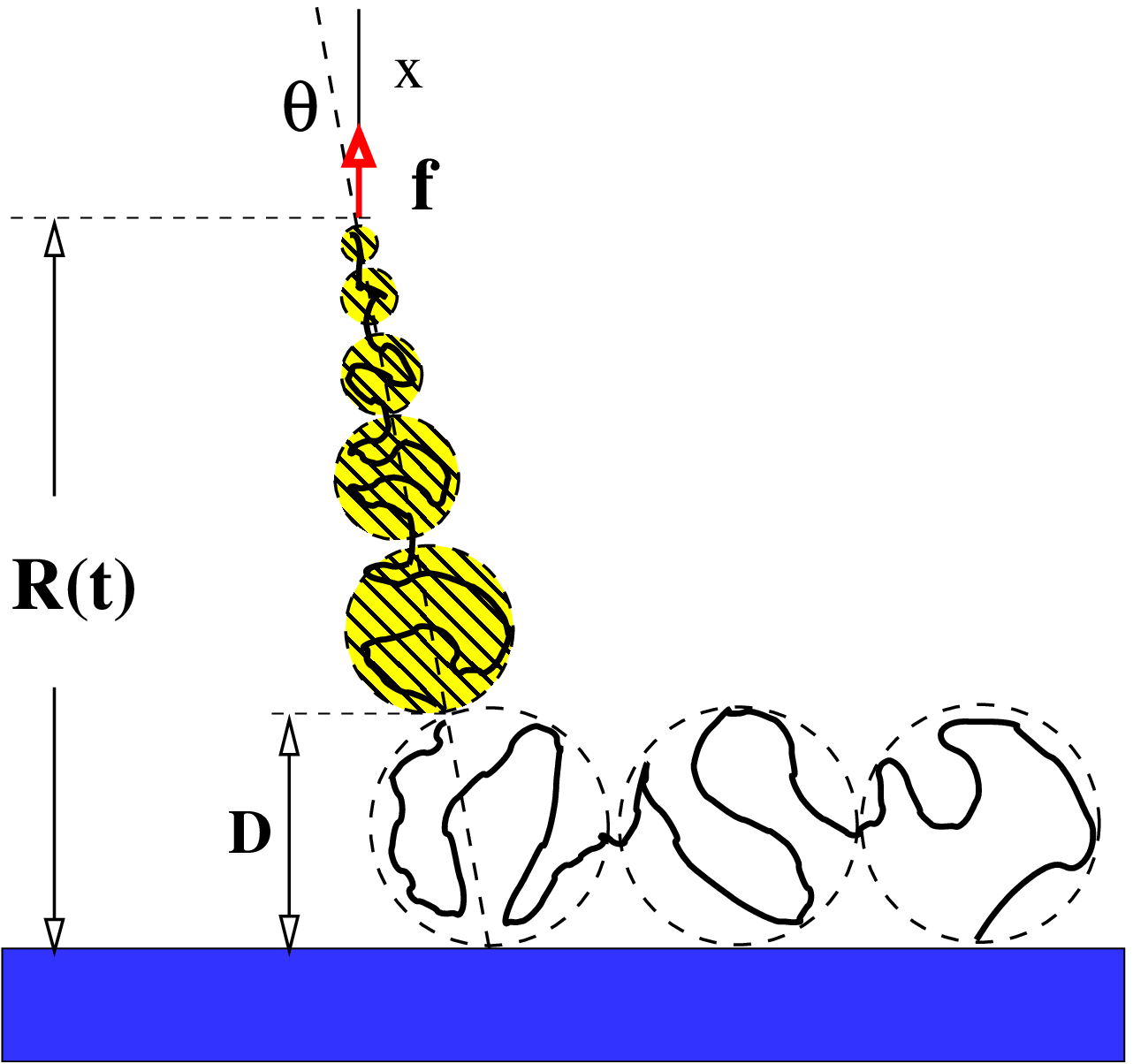}
\hspace{1cm}
\includegraphics[scale=0.35]{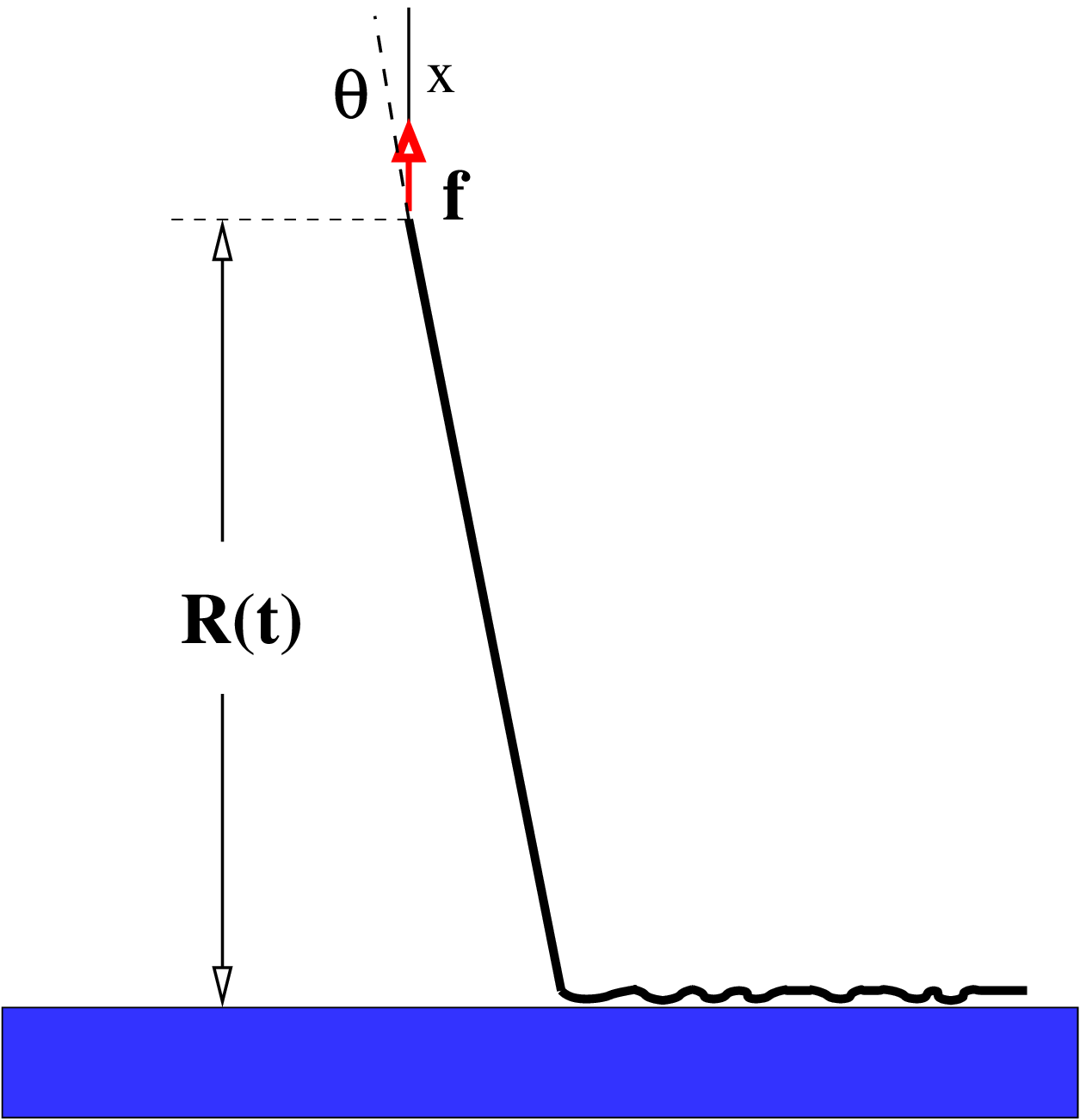}
\hspace{1cm}
\includegraphics[scale=0.35]{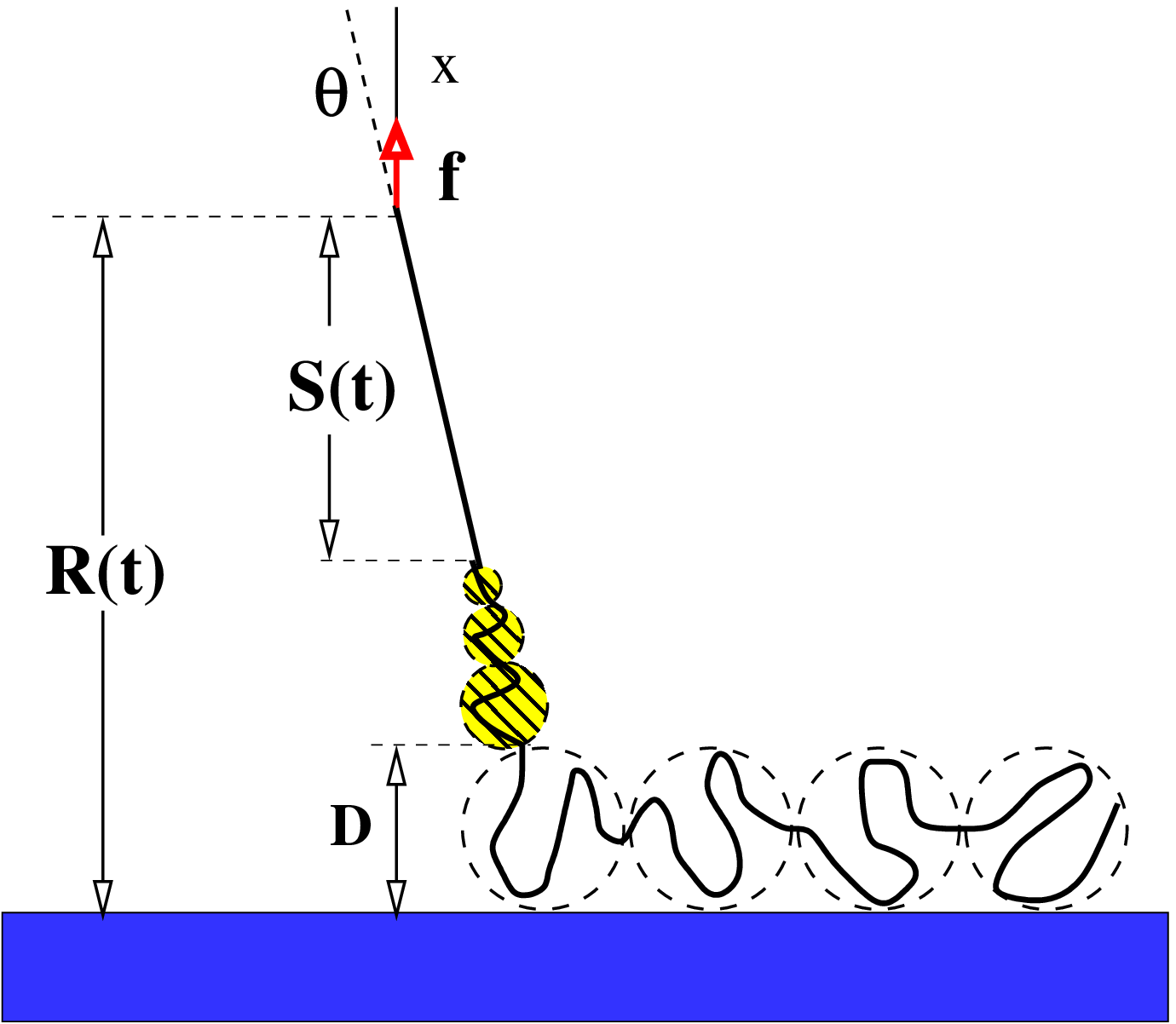}
\caption{The  picture of the single adsorbed chain detachment. The chain is
initially equilibrated, giving rise to a string of adsorption blobs with a blob
size $D$. The external driving force $f$, exerted on the first monomer, is
switched on at the initial time moment. By time $t$ the position of the first
monomer is denoted as $R(t)$. The shaded part of the chain displays a tensile
blobs sequence which has a form of a ``trumpet'', tilted opposite to the lateral
velocity direction by angle $\theta$. a) The ``trumpet'' regime, corresponding
to weak adsorption and weak pulling force, i.e., $1/N^{\nu} \ll {\widetilde
\varphi} < {\widetilde f} < 1$. b) The ``stem'' regime, corresponding to strong
adsorption and strong pulling force, $1 \ll {\widetilde \varphi} < {\widetilde
f}$. c) The ``Stem-trumpet'' regime, representing the intermediate pulling force
interval ${\widetilde \varphi} < 1 < {\widetilde f}$.}
\label{Blob}
\end{figure}

After equilibration on an attractive solid substrate, the adsorbed polymer chain
can be pictured as a two dimensional string of {\it adsorption blobs} of size $D
\propto  a (\epsilon_s - \epsilon_c)^{-\nu/\phi}$ \cite{Gennes}, where $a$ is
the Kuhn segment length, and $\nu$ and $\phi$ are  respectively the Flory
exponent and the crossover (or, adsorption) exponent. The dimensionless
adsorption energy $\epsilon_s = \varepsilon_s/k_BT$ (here and in what follows
$k_B$ denotes the Boltzmann constant, and $T$ is temperature), whereas
$\epsilon_c$ is the critical adsorption point (CAP). The external driving  force
$f$ acts on the free chain end (in the $x$-direction) while the second chain end
is tethered to the substrate. If the pulling force strength falls in the range
$1/N^{\nu} \ll a f /k_BT \leqslant 1$, the chain tail forms a kind of
``trumpet'' shape, as shown schematically in Fig. \ref{Blob}a. Moreover, as far
as the adsorption blob, adjacent to ``trumpet'', changes its lateral position
(i.e., its coordinates within the substrate) during desorption, the ``trumpet''
has not only vertical velocity (in the $x$-direction) but also a lateral one.
The latter leads to a tilt of the ``trumpet'' axis in direction opposite to the
lateral velocity due to Stokes friction. The tilting angle is denoted by
$\theta$ in Fig. ~\ref{Blob} and, as will be seen from our simulation results in
Sec. \ref{sec_simul}, almost does not change in the course of detachment. This
substantially simplifies the theoretical consideration.

Let us denote by $f(x, t)$ the tensile force which acts in direction of the
``trumpet'' axis at  an arbitrary cross-section of a tensile (Pincus) blob,
placed at distance $x$ from the substrate at time $t$. The $x$-component of this
tensile force reads $f_{x}(x, t) = f(x, t) \cos \theta$. The blob size is then
given by
\begin{eqnarray}
 \xi(x, t) = \dfrac{k_BT}{f(x, t)} = \dfrac{C k_BT }{f_{x}(x, t)}
\label{Blob_Size}
\end{eqnarray}
where $C = \cos \theta$.
It can be seen from Fig. \ref{Blob}a that the size of the tensile blob at $x$
corresponds to the lateral size of the polymer at height $x$. The balance of
forces in $x$-direction at this point yields
\begin{eqnarray}
 f_{x}(x, t) = \gamma \: \int\limits_{D}^{x} \: v_{x}(x', t) \: \left[
\dfrac{\xi(x', t)}{a}\right]^{z-2}
\dfrac{d x'}{\xi(x', t)} + \varphi
\label{Force_Balance}
\end{eqnarray}
where $v_{x}(x, t)$  is the $x$-component of the local velocity and $\varphi$ is
the {\it restoring force} which is exerted on the  moving domain by the last
adsorption blob. The value of $\varphi$  depends on the proximity to the
critical point $\epsilon_c$ as well as on the chain model itself (see below). In
the integral in Eq. (\ref{Force_Balance}), $dx'/\xi(x', t)$ counts the number of
blobs in the interval $x', x'+ dx'$ whereas $\gamma \: v_{x}(x', t)\: [\xi(x',
t)/a]^{z-2}$ is the local Stokes friction force ($\gamma$ being the friction
coefficient) \cite{Gennes_Book}. The dynamic exponent equals either $z = 2 +
1/\nu$, or $z = 3$, for Rouse- and Zimm dynamics respectively. Thus, taking into
account Eq.~(\ref{Blob_Size}), the equation for the blob size reads
\begin{eqnarray}
  \xi(x, t) = \dfrac{C k_BT}{\gamma \:  \int_{D}^{x} \: v_{x}(x', t) \:
\left[ \xi(x') / a \right]^{z-2} d x'/\xi(x') + \varphi}.
\end{eqnarray}
With this equation one may represent the local velocity $v(x, t)$ in terms of
the blob size $\xi (x, t)$ as
\begin{eqnarray}
 v_{x}(x, t) = - \dfrac{C k_BT} {a \gamma} \: \left[ \dfrac{a}{\xi(x, t)}
\right]^{z-1}\: \dfrac{\partial \xi(x, t)}{\partial x}.
\label{Local_Velocity_1}
\end{eqnarray}
The local (linear) monomer density $\rho(x, t)$ and the local blob size  $\xi(x,
t)$ are related by $\rho(x, t) = [\xi(x, t)/a]^{(1-\nu)/\nu}/a$ so that
Eq.~(\ref{Local_Velocity_1}) can be written in the form
\begin{eqnarray}
 v_{x}(x, t) = - \dfrac{C k_BT \nu}{\gamma (1 - \nu)} \:
\dfrac{1}{[{\widetilde \rho}(x, t)]^{\nu(z-2)/(1-\nu) + 1}} \: \dfrac{\partial
{\widetilde \rho}(x, t)} {\partial x},
\label{Velocity_2}
\end{eqnarray}
where we have introduced the notation for the dimensionless monomer density
${\widetilde \rho}(x, t) = a \rho (x, t)$.

The local flux of polymer beads has a standard form $J_{x}(x, t) = v_{x}(x, t)
\: \rho(x, t)$, that is,
\begin{eqnarray}
 J_{x}(x, t) = - \dfrac{ \nu C k_B T}{a\gamma (1-\nu) [{\widetilde \rho}(x,
t)]^{\nu (z-2)/(1-\nu)}}
\: \dfrac{\partial}{\partial x} \: {\widetilde \rho} (x, t)
\label{Flux}
\end{eqnarray}
With $J_{x}(x,t)$ from Eq. (\ref{Flux}), one may determine ${\widetilde
\rho}(x,t)$ from  the continuity equation $\partial \rho (x, t)/\partial t = -
\partial J_{x}(x, t) / \partial x$:
\begin{eqnarray}
 \dfrac{\partial {\widetilde \rho} }{\partial {\widetilde t}} \: =
\dfrac{\partial }{\partial {\widetilde x}} \left( {\widetilde \rho}^{m-1}
 \dfrac{\partial {\widetilde \rho} }{\partial {\widetilde x}}\right)
\label{PME}
\end{eqnarray}
where we have introduced the dimensionless (tilded) independent variables
${\widetilde x} = x/a$ and ${\widetilde t} = t/\tau_0$ (thereby the
characteristic time is $\tau_0 = a^2 \gamma (1-\nu) /\nu C k_B T$). Eq.
(\ref{PME}) has the form of the nonlinear {\it porous medium equation} (PME)
\cite{Vazquez} where  the  characteristic exponent $m = - [\nu (z - 1) - 1]/(1 -
\nu)$ has in our case a negative value. An equation similar to PME has been
discussed recently by Sakaue et al. \cite{Sakaue_3} in the context of polymer
expansion and unfolding from a compact state. In this case the consideration has
been based on the dynamics of {\it concentration blobs} \cite{Gennes_Book}
rather than tensile blobs like here. This kind of nonlinear partial differential
(parabolic) equations is commonly used in the context of groundwater
infiltration or seepage through porous strata \cite{Barenblatt} where the
exponent $m > 1$.

The geometric factor $C = \cos \theta$ is unessential in the subsequent
consideration and we will drop it. In addition to Eq.~(\ref{PME}), one has to
fix the initial and boundary conditions (BC). Initially, there are no polymer
segments at $x > D$, so that
\begin{eqnarray}
 {\widetilde \rho} (x, t = 0) = 0
\label{Initial}
\end{eqnarray}
At $x = D$ the tensile blob is equal in size to the adsorption blob, i.e., $\xi
(x = D, t) = k_BT/\varphi$. As a result, in terms of the dimensionless density,
one has
\begin{eqnarray}
 {\widetilde \rho} ( {\widetilde x} = {\widetilde D}, t) = \dfrac{1}{{\widetilde \varphi}^{(1-\nu)/\nu}}
\label{BC_1}
\end{eqnarray}
where ${\widetilde D} = D/a$ and ${\widetilde \varphi} = a \varphi/k_BT$.

The BC on the moving end of the ``trumpet'' follows from the condition $\xi (x =
R(t), t) = k_BT/f$ which corresponds to the smallest blob size, i.e.,
 \begin{eqnarray}
  {\widetilde \rho} ( {\widetilde x} = {\widetilde R}(t), t) = \dfrac{1}{{\widetilde f}^{(1-\nu)/\nu}}
 \label{BC_2}
\end{eqnarray}
Here again the dimensionless quantities, ${\widetilde R}(t)  = R(t)/a$ and
${\widetilde f} = a f/k_BT$, are introduced.

Our consideration is based on the assumption that the lateral component of the
tensile force is very weak (because the trumpet is almost vertical), so that any
force propagation along the adsorbed part of chain can be neglected. This is
different from the case considered by Serr and Netz \cite{Netz} where the
pulling of a strongly adsorbed chain by means of an AFM has been investigated.
In contrast to the present case where an uniform field pulls on the first bead,
the AFM cantilever can move vertically and horizontally with respect to
substrate so that the tilt angle $\theta$ changes with time. Moreover, in this
case the sliding of the polymer on the substrate becomes important with the
polymer-surface friction force  being much larger than the Stokes friction force
 for the detached part which is essential in our case. Therefore, the AFM can
not be actually used  for testing the scheme shown in Fig.  \ref{Blob}. In
Sec.~\ref{sec_exper} we will discuss a plausible electrostatic tracking
experiment which makes it possible, in principle, to test a gentle detachment
even close to the adsorption critical point.

\subsubsection{Self-similar solution}
\label{sss_self_similar}
In the same manner as for the PME, the solution of Eq. (\ref{PME}) is derived by
symmetry of  self-similarity (SS). This symmetry implies invariance of the
solution  with respect to the {\it stretching transformation} : ${\widetilde t}
\rightarrow \eta {\widetilde t} $ , ${\widetilde x} \rightarrow \eta^{\delta}
{\widetilde x} $, ${\widetilde \rho} \rightarrow \eta^{\omega} {\widetilde \rho}
$. The requirement that Eq. (\ref{PME}) stays invariant under the stretching
transformation connects the exponents $\omega, m$, and $\delta$ by the
following relation \cite{Vazquez}
\begin{eqnarray}
  \omega (1 - m)+2 \delta = 1
\label{Relation}
\end{eqnarray}
On the other hand, this requirement also fixes the self-similar form of the
solution as
\begin{eqnarray}
 {\widetilde \rho} ({\widetilde x}, {\widetilde t}) = t^{\omega} \: W(u), \quad
u = ({\widetilde x} - {\widetilde D})/{\widetilde t}^{\delta}.
\label{SS}
\end{eqnarray}
In Eq.~(\ref{SS}), $W(u)$ is a scaling function and we took into account that at
${\widetilde t} = 0$ ${\widetilde x} = {\widetilde D}$. On the other hand, the
boundary condition given by Eq. (\ref{BC_1}) is compatible with the SS-form Eq.
(\ref{SS}) only if $\omega = 0$. Then, as a consequence of Eq. (\ref{Relation}),
$\delta = 1/2$. Substituting  Eq. (\ref{SS}) into (\ref{PME}) yields a nonlinear
ordinary differential equation for the scaling function $W(u)$, i.e.,
\begin{eqnarray}
 \dfrac{d}{d u}\left(W^{m-1}  \dfrac{d}{d u}W\right)  + \dfrac{u}{2}
\dfrac{d}{d u}W  = 0
\label{ODE}
\end{eqnarray}
The BC given by Eq. (\ref{BC_1}) reads
\begin{eqnarray}
 W(u = 0) = \dfrac{1}{{\widetilde \varphi}^{(1-\nu)/\nu}}
\label{BC_1_New}
\end{eqnarray}

Assume now that the moving chain end follows the scaling relation ${\widetilde
R} ({\widetilde t}) = {\widetilde D} + {{\widetilde v}_d} {\widetilde t}^{\chi}$
where $\chi$ is an exponent and the factor ${\widetilde v}_d$ will be fixed
below. This scaling ansatz is compatible with the BC given by Eq. (\ref{BC_2})
only if $\chi = 1/2$. This imposes the  boundary condition
\begin{eqnarray}
 W(u = {\widetilde v}_d) = \dfrac{1}{{\widetilde f}^{(1-\nu)/\nu}}
\label{BC_2_New}
 \end{eqnarray}
Finally, the initial condition Eq. (\ref{Initial}) becomes
\begin{eqnarray}
 W(u = \infty) = 0
\label{Initial_1}
\end{eqnarray}

The nonlinear Eq. (\ref{ODE}) may be solved numerically. However, in case the
pulling force $f$ exceeds the restoring force only by a small amount, one can
show (see below) that ${\widetilde v}_d \ll 1$, i.e., the interval of argument
$u$ variation, where $W(u)$ is nonzero, is very narrow. As a result, $0\le u \le
{\widetilde v}_d \ll 1$ and one can linearize Eq. (\ref{ODE}) (by making use Eq.
(\ref{BC_1_New})) to get
\begin{eqnarray}
 {\widetilde \varphi}^{z-2} \dfrac{d^2}{d u^2} W(u) + \dfrac{u}{2} \dfrac{d}{d u} W(u) = 0
\label{ODE_Lin}
\end{eqnarray}
where  for small arguments $W(u)^{m-1} \approx {\widetilde \varphi}^{(z-2)}$.
The solution of Eq. (\ref{ODE_Lin}), subject to both conditions Eq.
(\ref{BC_1_New}) and Eq. (\ref{Initial_1}), is given by
\begin{eqnarray}
 W(u) = \dfrac{1}{{\widetilde \varphi}^{(1 - \nu)/\nu}}\left[ 1 -
{\rm erf} \left( \dfrac{u}{2 {\widetilde \varphi}^{(z - 2)/2}}\right) \right]
\end{eqnarray}
where ${\rm erf}(x)= (2/\sqrt{\pi}) \int_{0}^{x} \exp(-t^2) dt$ is the error
function.

The chain end velocity amplitude, ${\widetilde v}_d$, can be determined now from
the condition Eq. (\ref{BC_2_New}). If the pulling force ${\widetilde f}$ is
only slightly larger than the restoring force ${\widetilde \varphi}$, one can
expand $W(u) \approx  [1 - u/(\pi^{1/2} {\widetilde \varphi}^{(z-2) / 2})] /
{\widetilde \varphi}^{ (1-\nu)/\nu}$. Thus, one obtains
\begin{eqnarray}
 {\widetilde v}_d \simeq \sqrt{\pi} \: {\widetilde \varphi}^{(z - 2)/2}\left[ 1 -
\left( \dfrac{{\widetilde \varphi}}{{\widetilde f}}\right)^{(1 - \nu)/\nu}\right]
\label{Factor}
\end{eqnarray}

The equation of motion for the moving end then reads
\begin{eqnarray}
 {\widetilde R}({\widetilde t}) = {\widetilde D} + \sqrt{\pi}
{\widetilde \varphi}^{(z - 2)/2}\left[ 1 - \left( \dfrac{{\widetilde
\varphi}}{{\widetilde f}}\right)^{(1 - \nu)/\nu}\right] \sqrt{\widetilde t}
\label{R}
\end{eqnarray}

The variation of the number of desorbed monomers with time, $M(t)$, is given by
the integral
\begin{eqnarray}
 M(t) &=& \int\limits_{D}^{R(t)} \: \rho (x, t)\; dx =
\sqrt{{\widetilde t}} \: \int\limits_{0}^{{\widetilde v}_d} \; W(u) d u \nonumber\\
&\simeq&  \sqrt{\pi}{\widetilde \varphi}^{(z \nu - 2)/2\nu} \left[ 1 -
\left( \dfrac{{\widetilde \varphi}}{{\widetilde f}}\right)^{(1-\nu)/\nu} \right]
\sqrt{ {\widetilde t}}
\label{M}
\end{eqnarray}

Eventually, the total time for polymer detachment, $ \langle \tau_d \rangle$,
reads
\begin{eqnarray}
 \langle \tau_d \rangle = \dfrac{\tau_0 N^2}{\pi {\widetilde \varphi}^{(z \nu - 2)/2\nu}
\left[1 - ({\widetilde \varphi}/ {\widetilde f})^{(1-\nu)/\nu} \right]^2}
\label{N}
\end{eqnarray}
Combining Eqs. (\ref{M}) and (\ref{N}), one obtains the following dynamical
scaling law
\begin{eqnarray}
 M(t) = N \sqrt{\dfrac{{\widetilde t}}{\langle \tau_d \rangle}}
\label{Scaling_M}
\end{eqnarray}

Eventually, we would like to stress that in the foregoing consideration one
neglects the thermal fluctuations so that the necessary condition for desorption
requires that ${\widetilde f} > {\widetilde \varphi}$. On the other hand, the
``trumpet'' formation is only possible when $1/N^{\nu} \ll {\widetilde f} < 1$.
Therefore, the ``trumpet'' regime holds, provided that
\begin{eqnarray}
 1/N^{\nu} \ll {\widetilde \varphi} <{\widetilde f} < 1,
\end{eqnarray}
which means that then one deals with weak adsorption and weak pulling force.

\subsection {Dependence of the restoring force $\varphi$ on the adsorption
energy $\epsilon_s$}

One can write the restoring force in scaling form, $\varphi = \frac{k_BT}{a}
{\cal F}(\epsilon_s)$, with $\epsilon_s = \varepsilon_s/k_BT$ and ${\cal F}(x)$
- a scaling function. Force $\varphi$ stays constant in the course of the
desorption process as long as at least one adsorption blob is still located on
the attractive surface. The strength of $\varphi$ can be determined from the
plateau in the deformation curve: force vs. chain end position
\cite{Bhattacharya_1}.

In an earlier work~\cite{Bhattacharya_2}, we demonstrated that the fugacity  per
adsorbed monomer $z^{*}(\epsilon_s)$, which determines its chemical potential $
\mu_{\rm ads} = \ln z^{*}$,  can be found from the basic equation
\begin{eqnarray}
 \Phi (\alpha, \mu_3 z^*) \; \Phi (\lambda, \mu_2 w z^*) = 1.
\label{Basic}
\end{eqnarray}
In Eq.~(\ref{Basic}) $\alpha = \phi - 1$ ($\phi$ is the so-called crossover
exponent which, according to  the different simulation methods, ranges between
$\phi  \simeq 0.48$ and $\phi  \simeq 0.59$; see the corresponding discussion in
ref.~\cite{Bhattacharya_2} )  $\lambda = 1 - \gamma_{d=2} \simeq - 0.343$ (where
$\gamma_{d=2}$ is a universal exponent which governs the two-dimensional
polymer statistics), $w = \exp (\epsilon_s)$, $\mu_2$, and $\mu_3$ are so-called
{\it connective constants} in two- and three-dimensional spaces respectively,
and  the polylog function $\Phi$ is defined by the series  $\Phi(\alpha, y) =
\sum_{n=1}^{\infty} \: y^n / n^{\alpha}$. The values of $\mu_2$, $\mu_3$ and
$\gamma_{d=2}$ could be found in the text-book \cite{Vanderzande}.

In local equilibrium $ \mu_{\rm ads}$ should be equal to the chemical potential
of detached monomer $ \mu_{\rm det} = \dfrac{1}{k_B T} \; \dfrac{\partial F_{\rm
str}}{\partial N}$, where $F_{\rm str}$ is the free energy of the  {\it driven}
polymer chain so that $\mu_{\rm ads} = \mu_{\rm det}$. The expression for
$F_{\rm str}$ depends on the polymer model. The moving blob domain, shown in
Fig. \ref{Blob}, exerts on the last adsorbed monomer a force which is equal in
magnitude and opposite in sign to the restoring force $\varphi$. Therefore, the
free energy of a stretched polymer within the {\it bead-spring} (BS) model is
given by \cite{Bhattacharya_2}
\begin{eqnarray}
 F_{\rm str} = - N \left( \dfrac{a \varphi}{k_B T}\right)^{1/\nu} - N \ln \mu_3
\label{Free_BS}
\end{eqnarray}

On the other hand, within the freely jointed bond vector (FJBV) model, the
partition function, corresponding to $F_{\rm str}$,  reads \cite{Lai}
\begin{eqnarray}
 Z_{N} = \left( 4 \pi\right)^N \: \left( \dfrac{k_B T}{a \varphi}\right)^N \:
\left[ \sinh\left(\dfrac{a \varphi}{k_B T} \right) \right]^N
\label{FJBV_1}
\end{eqnarray}
In Eq. (\ref{FJBV_1})  the free chain partition function $Z_N = \left( 4 \pi
\right)^N$  for $\varphi \rightarrow 0$. A correspondence with the BS-model
may be established by the substitution $4 \pi \rightarrow \mu_3$ in Eq.
(\ref{FJBV_1}) so that the FJBV-free energy becomes
\begin{eqnarray}
 F_{\rm str} = - k_B T N \ln \left[ \mu_3 \left( \dfrac{k_B T}{a \varphi}\right)
\sinh \left(\dfrac{a \varphi}{k_B T} \right) \right]
\label{FJBV_Free}
\end{eqnarray}

Taking into account Eqs. (\ref{Free_BS}) and (\ref{FJBV_Free}), one obtains
eventually
\begin{eqnarray}
 \dfrac{a \varphi}{T} = \begin{cases}
                         \left[ - \ln (\mu_3  z^{*}(\epsilon_s)) \right]^{\nu} ,
&\mbox{BS model}\\
{\cal R}^{-1} (- \ln (\mu_3  z^{*}(\epsilon_s))), &\mbox{FJBV model}
                        \end{cases}
\label{Resistance_Force}
\end{eqnarray}
where ${\cal R}^{-1} (x)$ is the inversed of the function ${\cal R} (x) = \ln
\left[ \mu_3 \sinh (x)/x \right]$ and $\mu_3 z^{*}(\epsilon_s) \le 1$.


Below we consider separately the restoring force, Eq. (\ref{Resistance_Force}),
in the limits of strong and weak adsorption.

\subsubsection{Strong adsorption}

In the case of the strong adsorption ($\epsilon_s \gg 1$), the solution of Eq.
(\ref{Basic}) reads $z^{*}(\epsilon_s) \simeq \mu_2^{-1} {\rm e}^{-\epsilon_s}$
\cite{Bhattacharya_2}, and for the BS-model one gets $a \varphi/k_BT \simeq
\left[
\epsilon_s + \ln (\mu_2/\mu_3)\right]^{\nu}$.

For the FJBV model at $\epsilon_s \gg 1$ one has ${\cal R} (x) \simeq \ln \left(
\mu_3 {\rm e}^{x}/x\right) \simeq x + \ln \mu_3$. From Eq.
(\ref{Resistance_Force}) it then follows $a \varphi/k_B T + \ln \mu_3 =
\epsilon_s +
\ln \mu_2$, or $a \varphi/k_B T \simeq \epsilon_s + \ln (\mu_2/\mu_3$. The final
result for the restoring force in the strong adsorption limit takes on the form
\begin{eqnarray}
 {\widetilde \varphi} = \begin{cases}
\left[ \epsilon_s + \ln (\mu_2/\mu_3)\right]^{\nu} ,
&\mbox{BS model}\\
\epsilon_s + \ln (\mu_2/\mu_3), &\mbox{FJBV  model}
                        \end{cases}
\label{Strong_Ads_Final}
\end{eqnarray}
It is well known that under strong forces (and correspondingly strong
adsorption) the FJBV-model is better suited for description of the simulation
experiment results (see, e.g., Sec. 5 in our paper \cite{Bhattacharya_1}). That
is why in Eq. \ref{Strong_Ads_Final} the second line is appears as the more
appropriate result.

\subsubsection{Weak adsorption}

Close to the critical adsorption point, the solution of Eq. (\ref{Basic}) takes
on the form

$z^{*}(\epsilon_s) \simeq  \mu_3^{-1}\left[ 1 - C_1 (\epsilon_s -
\epsilon_c)^{1/\phi} \right]$ \cite{Bhattacharya_2}, where $C_1$ is a constant.
Then,  Eq. (\ref{Resistance_Force})  gives $a \varphi/k_B T \simeq C_1
(\epsilon_s -
\epsilon_c)^{\nu/\phi}$ for the BS-model.

In the  case of weak adsorption  $a \varphi/k_B T$ is small and ${\cal R} (x)
\simeq
\ln \left[ \mu_3 (1 + x^2/6)\right] \simeq \ln \mu_3 + x^2/6$. Eq.
(\ref{Resistance_Force}) then yields $a \varphi/k_B T \simeq B_1 (\epsilon_s -
\epsilon_c)^{1/2\phi}$, where $B_1$ is a constant. The unified result in the
weak adsorption limit is
\begin{eqnarray}
 {\widetilde \varphi} = \begin{cases}
C_1 \left( \epsilon_s - \epsilon_c\right)^{\nu/\phi} ,
&\mbox{BS model}\\
B_1 (\epsilon_s - \epsilon_c)^{1/2 \phi}, &\mbox{FJBV  model}
                        \end{cases}
\label{Weak_Ads_Final}
\end{eqnarray}

In fact, Eq. (\ref{Weak_Ads_Final}) may also be derived differently. Namely, as
already mentioned, the size of an adsorption blob is given by $D = a C_1
(\epsilon_s - \epsilon_c)^{- \nu/\phi}$ whereas the size of a tensile blob
$\xi(x=D) = k_B T/\varphi$. The  condition for detachment may be set as
$\xi(x=D) = D$ which leads again to Eq. (\ref{Weak_Ads_Final}) (whereby the
dimensionless notation ${\widetilde \varphi} = a \varphi/k_BT$ has been used).
On the other hand, the relationship between $D$ and $\varphi$, written as
$D\varphi/k_B T\simeq 1$, leads to the adsorption blob size for the FJBV-model:
$D \sim (\epsilon_s - \epsilon_c)^{-1/2\phi}$ (where the second line in Eq.
(\ref{Weak_Ads_Final}) has been used).

\subsection{Stem and stem-trumpet regimes}

For strong,  $ 1 < a \varphi/k_BT <a f/k_BT $, and moderate, $a \varphi/k_BT <
1 < a f/k_BT$, detachment force $f$ one has to consider the ``stem''(see
Fig. \ref{Blob}b) and ``stem-trumpet'' (see Fig. \ref{Blob}c) regimes,
respectively.

\subsubsection{Stem regime}

In this case the moving domain is build up from a ``stem'' of the length $M(t)$
(total number of desorbed monomers). The ``stem'' velocity is $d M(t)/d t$ (with
an accuracy of the geometrical factor $C = cos \theta$ which is not important
for scaling predictions),  so that the balance of driving and drag forces yields
 \begin{eqnarray}
  \gamma M(t) \: \dfrac{d M(t)}{d t} = f - \varphi
 \label{Stem_Equation}
\end{eqnarray}
 The solution of Eq. (\ref{Stem_Equation}) in the dimensionless units reads
\begin{eqnarray}
 M(t) = \sqrt{2({\widetilde f} - {\widetilde \varphi}) \: {\widetilde t}}
\label{Stem_Solution}
\end{eqnarray}
Evidently, this takes on the same form as Eq. (\ref{Scaling_M}),
$M(t) = N \sqrt{ {\widetilde t}/\langle {\widetilde \tau_d}
\rangle}$, whereby the mean detachment (dimensionless) time
$\langle {\widetilde \tau_d} \rangle = N^2 /[2 ({\widetilde f} - {\widetilde \varphi})]$.

\subsubsection{Stem-trumpet regime}

The ``stem-trumpet''  shape of the driven polymer is shown in  Fig. \ref{Blob}c.
In this case the local density  in the ``trumpet'' part  ${\widetilde
\rho}({\widetilde x}, {\widetilde t})$ is governed  by the same Eq. (\ref{PME})
but with a boundary condition in the junction point (i.e., in the point where
the ``stem'' goes over into the ``trumpet'' part). There, the blob size is $\xi
(x = R(t) - S(t), t) = a$. Thus, one gets
\begin{eqnarray}
 {\widetilde \rho} ({\widetilde x} ={\widetilde R} (t) - {\widetilde S}(t),
{\widetilde t}) = 1 \label{BC_Stem_Trumpet}
\end{eqnarray}
The boundary condition at ${\widetilde x} = {\widetilde D}$ has the same form,
Eq. (\ref{BC_1}), as before. The solution again has  the scaling form
${\widetilde \rho}({\widetilde x}, {\widetilde t}) = W(u)$. The moving end now
follows the law ${\widetilde R} (t) = {\widetilde D} + {\widetilde S}(t) +
{\widetilde u}_d \sqrt{{\widetilde t}}$, where the prefactor ${\widetilde u}_d$
can be fixed by using the condition $W(u = {\widetilde u}_d) = 1$. Consequently,
\begin{eqnarray}
 {\widetilde u}_d = \sqrt{\pi} {\widetilde \varphi}^{(z - 2)/2} \left[ 1-
{\widetilde \varphi}^{(1 - \nu)/\nu}\right]
\label{u_d}
\end{eqnarray}

In order to calculate the length of the ``stem'' portion, ${\widetilde S}(t)$,
one can use the results of the previous subsection. The only difference lies in
the fact that now the ``stem - trumpet'' junction point moves itself with
velocity ${\widetilde u}_d/(2\sqrt{{\widetilde t}})$ and the driving force is
${\widetilde f} - 1$. Therefore, the balance of driving and drag forces leads to
the following equation for ${\widetilde S}(t)$
\begin{eqnarray}
 \left( \dfrac{{\widetilde u}_d}{2\sqrt{{\widetilde t}}} + \dfrac{d}
{ d {\widetilde t}} \: {\widetilde S}(t) \right) \: {\widetilde S}(t) = {\widetilde f}
- 1
\label{Equation_S}
\end{eqnarray}
The solution of Eq. (\ref{Equation_S}) reads
\begin{eqnarray}
 {\widetilde S}(t) = \left[ \sqrt{\dfrac{{\widetilde u}_d^2}{4}  +
2 ({\widetilde f} -1 )} - \dfrac{{\widetilde u}_d}{2}\right] \:
\sqrt{{\widetilde t}}
\label{Solution_S}
\end{eqnarray}
The resulting equation for the average height of the end monomer becomes
\begin{eqnarray}
 {\widetilde R} (t) = {\widetilde D} +
\left[ \sqrt{\dfrac{{\widetilde u}_d^2}{4}  +
2 ({\widetilde f} -1 )} + \dfrac{{\widetilde u}_d}{2}\right] \:
\sqrt{{\widetilde t}}
\end{eqnarray}

The number of desorbed monomer can be easily calculated as an integral over the
the number of monomers in the ``trumpet'' part and in the ``stem'' portion,
i.e.,
\begin{eqnarray}
 M({\widetilde t})  &=&  \sqrt{{\widetilde t}} \: \int\limits_{0}^{{\widetilde u}_d} \: W(u) d u
 + {\widetilde S}({\widetilde t}) \nonumber\\
&=&  \left[ \sqrt{\dfrac{{\widetilde u}_d^2}{4}  +
2 ({\widetilde f} -1 )} -  \dfrac{{\widetilde u}_d }{2}
\left( 1 - \dfrac{2}{{\widetilde \varphi}^{(1-\nu)/\nu}} \right)\right] \: \sqrt{{\widetilde t}}
\label{Desorbed_Monomers}
\end{eqnarray}
where ${\widetilde u}_d$ is given by Eq. (\ref{u_d}). Evidently, all
characteristic scales as well as the number of desorbed monomers vary in time as
$\sqrt{t}$. This shows that the average desorption time $\langle \tau \rangle$
is proportional to $N^2$  as before.

\subsection{Dynamic Regimes}

\begin{figure}[ht]
\begin{center}
\includegraphics[scale=0.6]{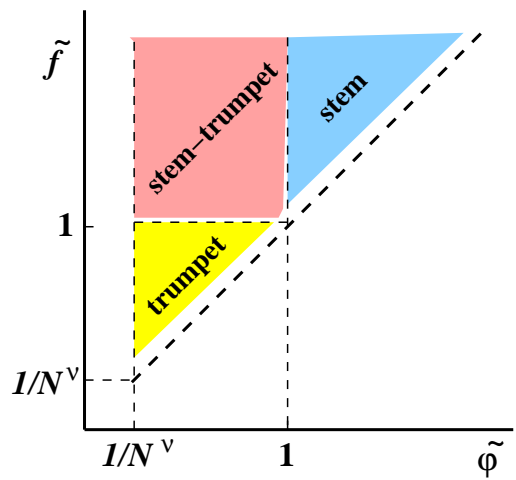}
\caption{Different dynamic regimes of forced polymer chain desorption in terms
of pulling ${\widetilde f}$ and restoring ${\widetilde \varphi}$ forces. The
relation between ${\widetilde \varphi}$ and the adsorption energy $\epsilon_s$
is given by Eq.~(\ref{Strong_Ads_Final}) and Eq.~(\ref{Weak_Ads_Final}) for the
strong and weak adsorption, respectively. The white strip along the diagonal
indicates that in all cases the condition ${\widetilde f} > {\widetilde
\varphi}$ is important. }\label{Diagram}
\end{center}
\end{figure}

The foregoing theoretical analysis was based essentially on the notion of
tensile (Pincus) blobs as well as on the ``quasi-static'' approximation, i.e.,
on the equating of driving and drag forces locally (in space and time).
Depending on the pulling  ${\widetilde f}$  and restoring ${\widetilde \varphi}$
force strengths, one distinguishes a ``trumpet'' ($1/N^\nu \ll  {\widetilde
\varphi} <  {\widetilde f} < 1$), ``stem-trumpet'' (${\widetilde \varphi} < 1 <
 {\widetilde f} $ ), and a ''stem`` ($1 <  {\widetilde \varphi} < {\widetilde
f}$) regimes (see Fig. \ref{Diagram}). Notably, in all regimes the average
number of desorbed monomers $M(t)$ and the height of the first monomer $R(t)$
evolve  as a $\sqrt{t}$-time universal law. Moreover, the evolution equation
which governs the polymer beads density $\rho (x, t)$ in the tensile blobs is
analogous to the nonlinear porous medium equation which is commonly used for
investigations of gas and fluid infiltration  in the geological porous strata
\cite{Vazquez,Barenblatt}. Following this line we have demonstrated that the
scaling analysis of such equation, based on symmetry of self-similarity,
immediately brings about a $\sqrt{t}$-time universal law.

The value of ${\widetilde \varphi}$ depends on the adsorption energy
$\epsilon_s$ and is described by different expressions, depending on the
particular polymer chain model: bead-spring (BC), or freely jointed bond vectors
(FJBV) models. For large adsorption energy ($\epsilon_s \gg 1$),  the restoring
force is also large, leading to the ''stem`` regime. Close to the the critical
adsorption point (for vanishing adhesion), the restoring force could attain a
very small value, leading to a `trumpet'', or ``stem-trumpet'' regimes. In this
case the role of fluctuations, which was neglected in our theoretical
investigation, becomes important and could result in correction to the basic
$\sqrt{t}$-time law.

In the next section we will give results of our MC- and MD-simulation study that
were carried out in some of the foregoing forced-desorption (mainly ''stem``)
regimes.

\section{Computational Models}
\label{sec_Model}
\subsection{Monte-Carlo}
\label{subsec_MC}

We have used a coarse grained off-lattice bead spring model to describe the
polymer chains. Our system consists of a single chain tethered at one end to a
flat structureless surface while the external pulling force is applied to other
chain end. In the MC simulation the surface attraction of the monomers is
described by a square well potential $U_w(x) = - \epsilon_s $ for $x<\delta =
0.125$  and $U_w(x) = 0$ otherwise. Here the strength $\epsilon_s /k_BT$ is
varied from $0.6$ to $3.6$. The effective bonded interaction is described by the
FENE (finitely extensible nonlinear elastic) potential.
\begin{equation}
U_{\mbox{\tiny FENE}}= -K(1-l_0)^2\ln\left[1-\left(\frac{l-l_0}{l_{max}-l_0}
\right)^2 \right]
\label{fene}
\end{equation}
with $K=20, l_{max}=1, l_0 =0.7, l_{min} =0.4$

The non-bonded interactions between the monomers are described by the Morse
potential.
\begin{equation}
\frac{U_M(r)}{\epsilon_M} =\exp(-2\alpha(r-r_{min}))-2\exp(-\alpha(r-r_{min}))
\end{equation}
with $\alpha =24,\; r_{min}=0.8,\; \epsilon_M/k_BT=1$. This model is currently
well established and has been used due to its reliability and efficiency in a
number of studies \cite{AMKB} of polymer behavior on surfaces.

\begin{figure}[htb]
\begin{center}
\includegraphics[scale=0.4]{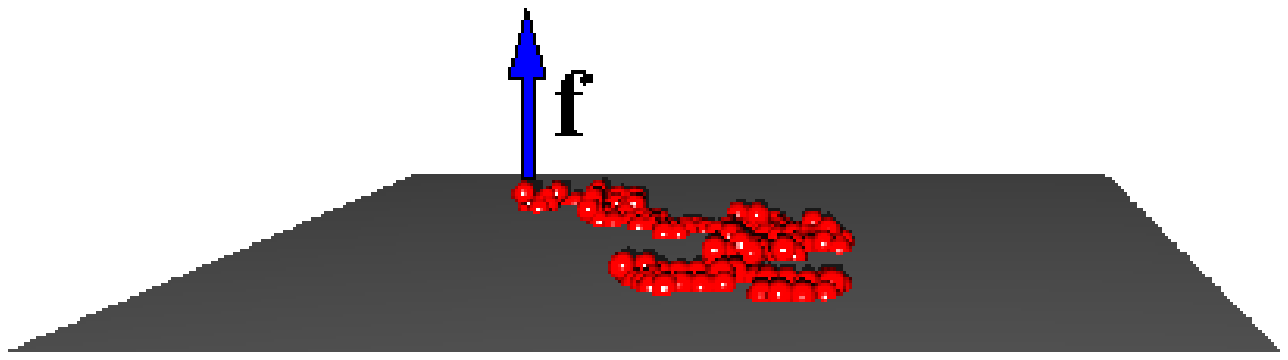}
\hspace{0.7cm}
\includegraphics[scale=0.4]{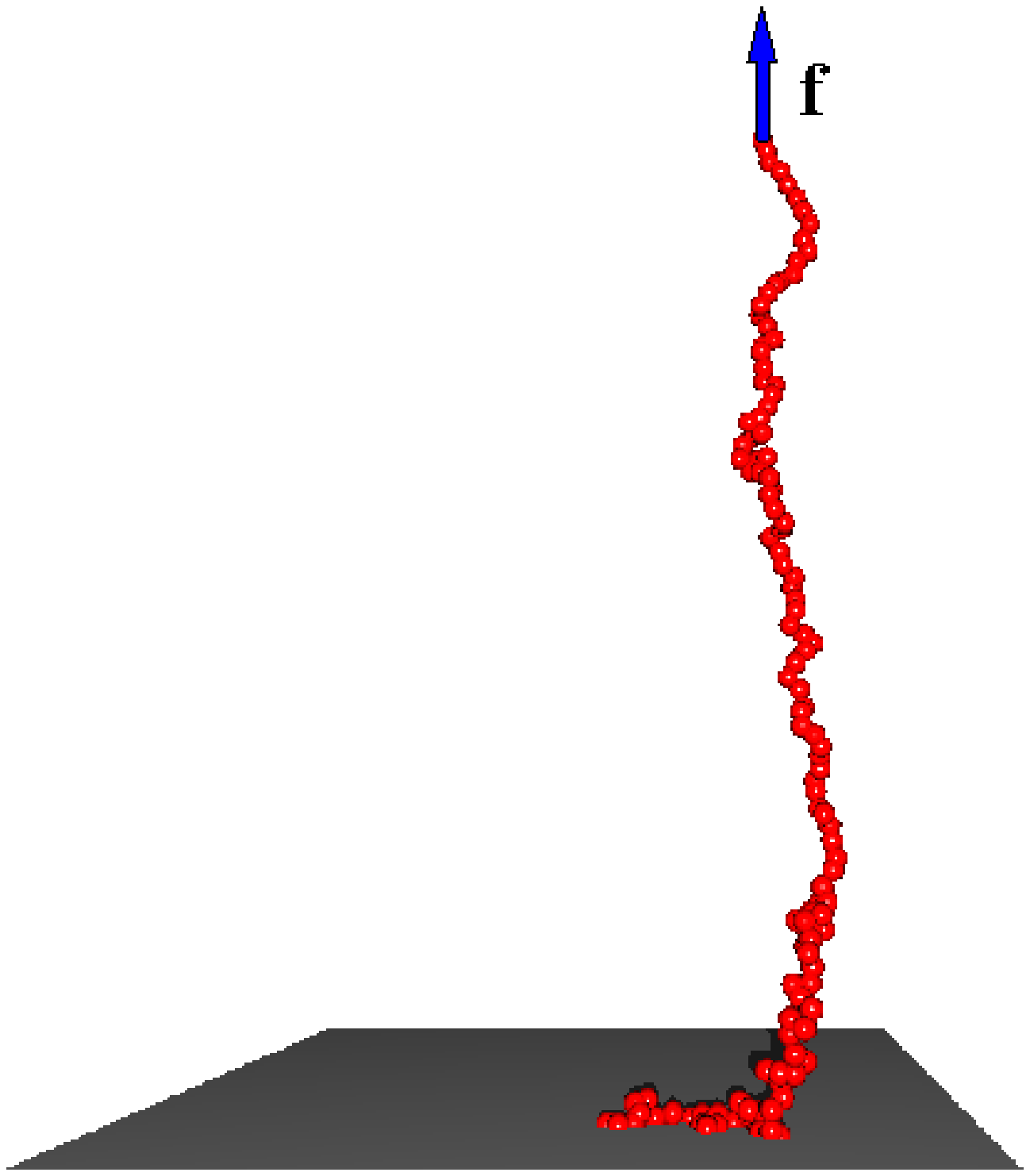}
\caption{Snapshots from a MC simulation of chain with $N = 128$, adsorbed on a
solid plane with energy of adsorption $\epsilon_s/k_BT = 3.0$. (a) at time $t =
0$ before, and (b) - after a pulling force $f=3.0$ is applied perpendicular to
substrate.}\label{fig_snapshot}
\end{center}
\end{figure}

We employ periodic boundary conditions in the $y-z$ directions and impenetrable
walls in the $z$ direction. The lengths of the studied polymer chains are
typically  $64$,  $128$, and $256$. The size of the simulation box was chosen
appropriately to the chain length, so for example, for a chain length of $128$,
the box size was $256 \times 256 \times 256$ . All simulations were carried out
for {\em constant} force. A force $f$ was applied to the last  monomer in the
$x$-direction, i.e., perpendicular to the adsorbing surface

The standard Metropolis algorithm was employed to govern the moves with  self
avoidance automatically incorporated in the potentials. In each Monte Carlo
update, a monomer was chosen at random and a random displacement was attempted
with $\Delta x,\;\Delta y,\;\Delta z$ chosen uniformly from the interval
$-0.5\le \Delta x,\Delta y,\Delta z\le 0.5$. The transition probability for the
attempted move was calculated from the change $\Delta U$ of the potential
energies before and after the move as $W=\exp(-\Delta U/k_BT)$. As for standard
Metropolis algorithm, the attempted move was accepted, if $W$ exceeds a random
number uniformly distributed in the interval $[0,1]$. As a rule, the polymer
chains have been originally equilibrated in the MC method for a period of about
$10^6$ MCS (depending on degree of adsorption $\epsilon_s$ and chain length $N$
this period is varied) whereupon one performs $200$ measurement runs, each of
length $8\times 10^6$ MCS. Various properties of the chain are then sampled
during the course of the run. The starting configuration is replaced by a new
sequence in the beginning of the next run. Two typical snapshots of chain
configurations, immediately before the detachment starts, and some time later,
are presented in Fig. \ref{fig_snapshot}.

\subsection{Molecular Dynamics}

In our MD-simulations we use  a coarse-grained model of a polymer chain of $N$
beads connected by finitely extendable elastic bonds. The bonded interactions in
the chain is described by the frequently used Kremer-Grest potential,
$V^{\mbox{\tiny KG}}(r)=V^{\mbox{\tiny FENE}}(r) + V^{\mbox{\tiny WCA}}(r)$,
with FENE potential, Eq.~(\ref{fene}), and a non-bonded repulsion term taken as
Weeks-Chandler-Anderson (WCA) (i.e., the shifted and truncated repulsive branch
of the Lennard-Jones potential) given by \begin{equation}
 V^{\mbox{\tiny WCA}}(r) = 4\epsilon\left[
(\sigma/ r)^{12} - (\sigma /r)^6 + 1/4
\right]\theta(2^{1/6}\sigma-r)
\label{wca}
\end{equation}
with $\theta(x)=0$ or 1 for $x<0$ or $x\geq 0$, and $\epsilon=1$, $\sigma=1$.
The potential
$V^{\mbox{\tiny KG}}(r)$ has got a minimum at bond length
$r_{\mbox{\tiny bond}}\approx 0.96$.

The substrate in the present investigation is considered simply as a
structureless adsorbing plane, with a Lennard-Jones potential acting with
strength $\epsilon_s$ in the perpendicular $x$--direction, $V^{\mbox{\tiny
LJ}}(x)=4\epsilon_s[(\sigma/x)^{12} - (\sigma/x)^6]$.

The dynamics of the chain is obtain by solving a Langevin equation of motion for
the position $\mathbf r_n=[x_n,y_n,z_n]$ of each bead in the chain,
\begin{equation}
m\ddot{\mathbf r}_n = \mathbf F_n^{j} + \mathbf F_n^{\mbox{\tiny
WCA}} -\gamma\dot{\mathbf r}_n + \mathbf R_n(t) +\mathbf f_n\delta_{nN}, \qquad
(n,\ldots,N)
\label{langevin}
\end{equation}
which describes the Brownian motion of a set of bonded particles whereby the
last of them is subjected to external (constant) stretching force $\mathbf f =
[f,0,0]$. The influence of solvent is split into slowly evolving viscous force
and rapidly fluctuating stochastic force. The random, Gaussian force $\mathbf
R_n$ is related to friction coefficient $\gamma$ by the fluctuation-dissipation
theorem. The integration step is $0.002$ time units (t.u.) and time in measured
in units of $\sqrt{m\sigma^2/\epsilon}$, where $m$ denotes the mass of the
beads, $m=1$. The ratio of the inertial forces over the friction forces in
Eq. (\ref{langevin}) is characterized  by the Reynolds number ${\rm Re} =
\sqrt{m \epsilon}/\gamma \sigma$ which in our simulation falls in the range
${\rm Re}= 0.1 \div 4$. In  the course of simulation velocity-Verlet algorithm
is used to integrate equations of motion (\ref{langevin}).

\section{Simulation results}
\label{sec_simul}

In order to check the validity of our theoretical predictions, we carried out
extensive computer simulations by means of both  Monte-Carlo (MC) and Molecular
Dynamics (MD) so that a comparison can be made between the overdamped dynamics
of a polymer (MC) and the behavior of an inertial (underdamped) system (MD). For
ultimate consistency of the data, we performed also MD simulations with very
large friction coefficient $\gamma = 10$ whereby the obtained data was found to
match that from the MC computer experiment.

\begin{figure}[ht]
\begin{center}
\includegraphics[scale=0.7]{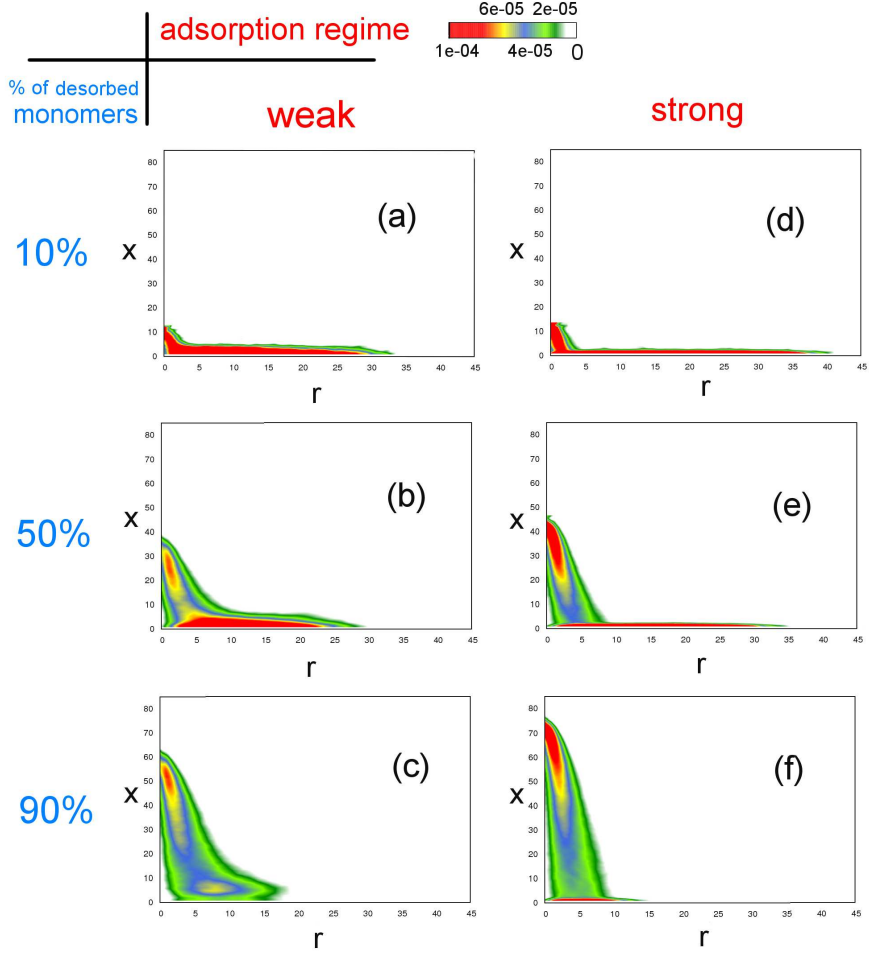}
\caption{Polymer  density distributions  of beads  plotted for different
adsorption regimes: (a)--(c) $\epsilon_s/(k_BT)=1.9$, (d)--(f)
$\epsilon_s/(k_BT)=5.0$ and various percentage of detached monomers ($10\%$,
$50\%$ and $90\%$). The $x$-axis coincides with  direction of pulling force
whereas $r=\sqrt{(y_1-y_n)^2+(z_1-z_n)^2}$, $(n=1,\ldots,N)$ is the radial
component of beads calculated with respect to the first monomer. Here $N=100$,
$T=0.1$ and $\gamma=0.25$.} \label{Density_Plot}
\end{center}
\end{figure}

First of all, it would be interesting to verify that trumpet-like  structures,
which have been discussed thoroughly in Sec.\ref{sec_theory}, can be seen in our
MD-simulation experiment. Figure \ref{Density_Plot} represents such a
visualization where the beads density distributions are given (as a function of
$x$-coordinate as well as the radial coordinate calculated with respect to the
first monomer, i.e. $r = \sqrt{(y_1-y)^2+(z_1-z)^2}$) for the weak ,
$\epsilon_s/k_B T = 1.9$, (left panel) and strong, $\epsilon_s/k_B T = 5$,
(right panel) adsorptions. Each density plot was implemented as a result of
averages over  many runs under the fixed fraction of desorbed monomers: $10\%$
(upper row), $50\%$ (middle row) and $90\%$ (bottom row). As one can see, the
moving domain is developed into the trumpet-like configuration in both, weak and
strong adsorption, cases. Moreover, the ''trumpet`` is tilted as it was argued
in Sec. \ref{sss_density} due to its lateral velocity and the corresponding
Stokes friction. An important point is that the tilt angle stays constant which
substantially simplifies the theoretical consideration in Sec.
\ref{sss_density}. Recall that the density in the blob $n_b(x)$ and the blob
size $\xi (x)$ are related as $n_b(x) \propto 1/[\xi (x)]^{3-1/\nu} \propto
1/[\xi (x)]^{1.3}$, i.e., the  larger the density is, the smaller blob size.
This tendency can be seen in Fig. \ref{Density_Plot} where the most dense
portion (red color) comes close to the tip of the ''trumpet`` as it should be.
Moreover, for the strong adsorption case (see Fig. \ref{Density_Plot}d-f) this
most dense portion becomes more extended which might mean a ''stem-trumpet``
formation.  As a result, the density plots, given in Fig.~\ref{Density_Plot},
provide a good evidence for the existence of the trumpet-like structures,
suggested and discussed in Sec.  \ref{sec_theory}.

\begin{figure}[ht]
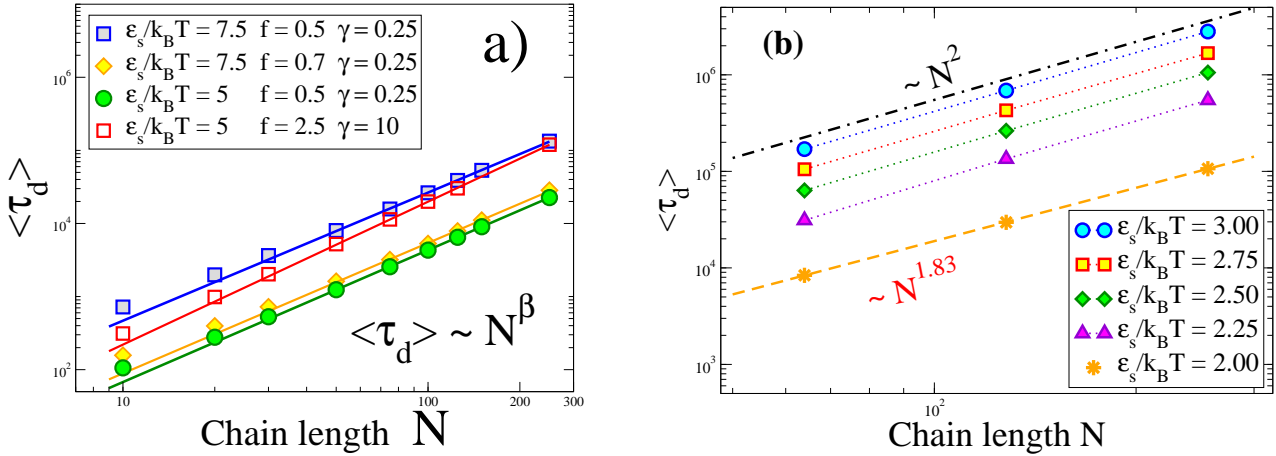

\vspace*{0.9cm}
\includegraphics[scale=0.3]{tauVSN.eps}
\hspace{0.7cm}
\includegraphics[scale=0.34]{tau_N_MC.eps}
\caption{(a) MD data for the mean desorption time $\langle \tau_d \rangle$ of a
chain vs chain length $N$, plotted for different strength of pulling force $f$,
adhesion strength $\epsilon_s/k_BT$, and friction $\gamma$. Symbols denote the
results of simulation while solid lines represent fitting curves $\langle
\tau_d\rangle \propto N^{\beta}$ which were found in the range $N=50 \div 250$.
The value of the exponent $\beta$ depends on the dynamics: underdamped
$\beta\approx 1.75\pm0.1$ (full symbols) and overdamped $1.96\pm0.03$ (empty
symbols). (b) Variation of $\langle\tau_d\rangle$ with $N$ at different strength
of adhesion $\epsilon_s/k_B T$ from the MC data. Note that here the critical
adsorption energy $\epsilon_c /k_B T \approx 2.0$. In the strong adsorption
regime $\beta = 2.0$, whereas close to the critical point $\beta = 1.83$. }
\label{fig_tauVSN}
\end{figure}

As one of our principal results, it was found that the chain length dependence
of the mean detachment time $\langle \tau_d \rangle$ follows a scaling law
$\langle \tau_d \rangle \sim N^{\beta}$, Fig. \ref{fig_tauVSN}, where $\beta$,
the detachment exponent, depends on the type of dynamics (underdamped or
overdamped) as well as on the adsorption strength (i.e., on the proximity of
$\epsilon_s$  to the adsorption critical point, $\epsilon_c$).
Fig.~\ref{fig_tauVSN}a shows result from the MD-simulation for different values
of $\epsilon_s$, the pulling force $f$, and the friction coefficient $\gamma$.
For the underdamped dynamics, one finds a detachment exponent $\beta = 1.75\pm
0.1$ whereas for the overdamped one $\beta = 1.96\pm 0.03$. The last value is
very close to the theoretical prediction which immediately follows from Eq.
(\ref{Stem_Solution}), $\langle \tau_d \rangle \propto \tau_0 N^2/ ({\widetilde
f} - {\widetilde \varphi}) $, i.e., $\beta = 2$. Recall that this theoretical
prediction neglects any effects of inertia and describes overdamped dynamics. In
addition, detachments are considered in  the ''stem`` scenario, that is, for
relatively large adsorption energy and pulling force.

This conclusion is supported by the MC-simulation results (which represent
the overdamped limit)  shown in Fig.  \ref{fig_tauVSN}b. For the relatively
strong adsorption one finds $\beta = 2.02 \pm 0.01$, which is very  close to the
theoretical prediction. On the other hand, in the vicinity of the adsorption
critical point the detachment exponent appears to be smaller $\beta = 1.83\pm
0.01$. Indeed, one may suggest that the observed decrease is due to
fluctuations which become stronger close to $\epsilon_c$. As mentioned above,
fluctuations are not taken into account in our theoretical treatment which
is based  on the ''quasistatic`` approximation. As a matter of fact, in the
cases of ''trumpet`` and ''stem-trumpet`` scenarios (relatively weak adsorption
energy and pulling force), the fluctuations alter the value of the detachment
exponent so that $\beta < 2$. It is, however, conceivable too that $\beta =
1.83$ may also be due to finite-size effects in the vicinity of the critical
adsorption point where the size of adsorption blobs becomes large.

\begin{figure}[ht]
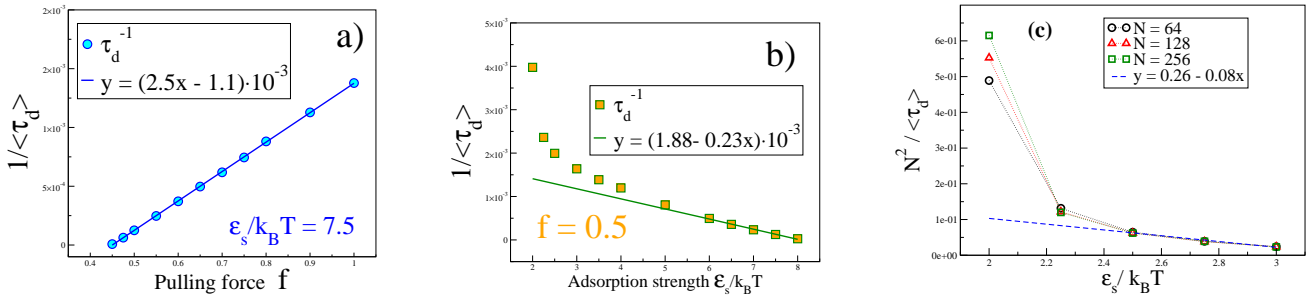

\includegraphics[scale=0.19]{tauVSf.eps}
\hspace{0.7cm}
\includegraphics[scale=0.19]{tauVSeps.eps}
\hspace{0.7cm}
\includegraphics[scale=0.22]{tau_E_MC.eps}
\caption{MD data on the inverse  mean desorption time $1/\langle\tau_d\rangle$
of a chain with $N=50$ plotted as a function of: a) pulling force $f$, and b)
adsorption strength $\epsilon_s/k_BT$. Here $\gamma=0.25$. c) Inverse mean
detachment time $N^2 / \langle \tau_d \rangle$ (scaled with $N^2$) against
$\epsilon_s$ from MC data with $f=3.0$ and $\epsilon_s/k_BT=3.0$.}
\label{fig_tauVSf}
\end{figure}

In order to check the $f-$ and $\epsilon_s-$dependence of
$\langle\tau_d\rangle$,
which the relation $\langle \tau_d \rangle \propto \tau_0 N^2/ ({\widetilde f} -
{\widetilde \varphi}) $ predicts for the the ''stem`` regime, we have plotted
$1/\langle \tau_d \rangle$ vs. $f$, Fig.~\ref{fig_tauVSf}a, and vs.
$\epsilon_s$, Fig.~\ref{fig_tauVSf}b. In Fig.~ \ref{fig_tauVSf}a one can see a
perfect straight line as expected. As for the $\epsilon_s$-dependence, we recall
that according to Eq.~(\ref{Strong_Ads_Final}),  in the strong adsorption limit
the restoring force ${\widetilde \varphi} \approx c_1 (\epsilon_s + c_2)$ where
$c_1$ and $c_2$ are some constants. As can be seen from Fig.~\ref{fig_tauVSf}b,
in the MD-simulation this holds for $\epsilon_s > 5$. On the other hand, in the
MC simulation, the strong adsorption regime starts at $\epsilon_s > 2.5$ as is
evident from Fig.~\ref{fig_tauVSf}c.

\begin{figure}[ht]
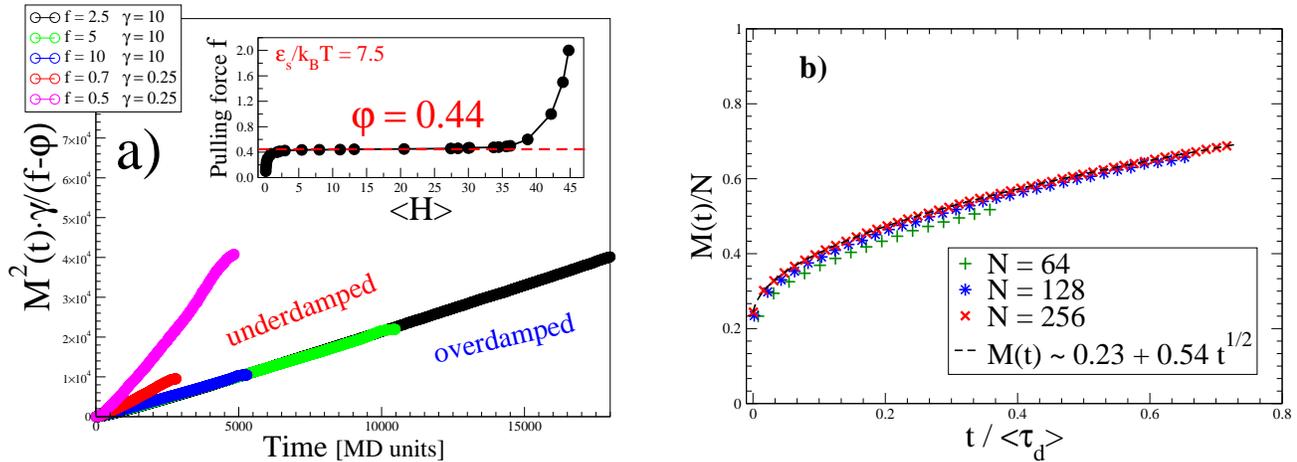

\vspace{0.5cm}
\includegraphics[scale=0.3]{fraction.eps}
\hspace{0.7cm}
\includegraphics[scale=0.33]{op_t_scal.eps}
\caption{ (a) Mean squared number of desorbed monomers $\gamma M^2(t) / [(f -
\varphi) ]$ for a chain composed of $N=100$ particles shown as a function of
time for different sets of $f$ and $\gamma$ - MD data. 
Here $\epsilon_s/k_BT=5.0$. 
The inset displays the
(equilibrium) relationship between pulling force $f$ and the resulting mean
height $\langle H \rangle$ of the first monomer in a chain with $N=50$. The
estimated value of the plateau height (which is the restoring force in the
course of detachment) is $\varphi = 0.44$. (b)
Variation of $M(t)$ with (dimensionless) elapsed time (measured in units of
$t/\langle \tau_d \rangle$) for different polymer lengths $N$ from MC data.
Here $\epsilon_s/k_BT=3.0$. Dashed line denotes the expected
$\sqrt{t}$-behavior, cf.
Eq.~(\ref{Stem_Solution}).}
\label{fig_OP}
\end{figure}

Fig.~\ref{fig_OP} demonstrates the validity of the $\sqrt{t}$-law for the time
evolution of the mean number of desorbed monomers, $M(t)$, given by
Eq.~(\ref{Stem_Solution}). According to this relationship, the quantity $\gamma
M^2(t)/(f - \varphi)$ should be a linear function of time.

In order to fix the value of the restoring force $\varphi$, we have carried out
an independent simulation to find how the mean height of the first monomer
$\langle H \rangle$ depends in equilibrium on the pulling force $f$ for a
previously adsorbed (end-grafted) chain. The result of this is shown as an
inset in Fig. \ref{fig_OP}a  where the value of plateau height yields the
restoring force $\varphi = 0.44$ at adsorption energy $\epsilon_s = 5.0$. By
making use of this value we were able to superimpose three curves for different
forces, $f=2.5,\; 5,\; 10$ in  the overdamped case ($\gamma = 10.0$) on a single
master curve as  shown in Fig. \ref{fig_OP}a. Evidently, in the case of
underdamped dynamics this collapse on a single master curve does not work. This
deviation from the theoretical predictions indicates again that they apply
perfectly only to the case of overdamped dynamics. Finally, the respective
$M(t)$ vs. time relationship from the MC-simulation is shown in Fig.
\ref{fig_OP}b. One should thereby note that in the starting equilibrium
configuration of the adsorbed chain about 24\% of the monomers reside in loops
so they are actually desorbed even before the pulling force is applied.
Evidently, as in the MD data for very large friction, the curves for different
chain lengths collapse on a single master $\sqrt{t}$-curve in agreement with
Eq.~(\ref{Stem_Solution}).

\section{Plausible experimental validation}
\label{sec_exper}

Now we are in a position to discuss how the detachment kinetics considered in
this paper could be validated in a laboratory experiment. To this end we recall
that using conventional AFM, optic (OT), or magnetic tweezers (MT), the tracking
is performed by practically macroscopic, of micrometers large, agents: by
cantilever in AFM, or by e.g. polystyrene beads in OT and MT cases. On the other
hand, the trumpet-like structures are the result of delicate interplay between
pulling and drag (or friction) forces. Obviously, the Stokes friction of the
cantilever as that of the polystyrene beads exceeds largely the friction of the
tested polymer chain itself and thus suppresses the observation of the predicted
time-behavior. One could solve  this problem only if the the tracking agent is
of the size of a single monomer, e.g., when it is a charged ionic end-group like
a carboxylate group placed in an external electric field. Moreover, the polymer
end should be fluorescently labeled in order to measure the polymer end spatial
location.

\begin{figure}[ht]
\begin{center}
\includegraphics[scale=0.4]{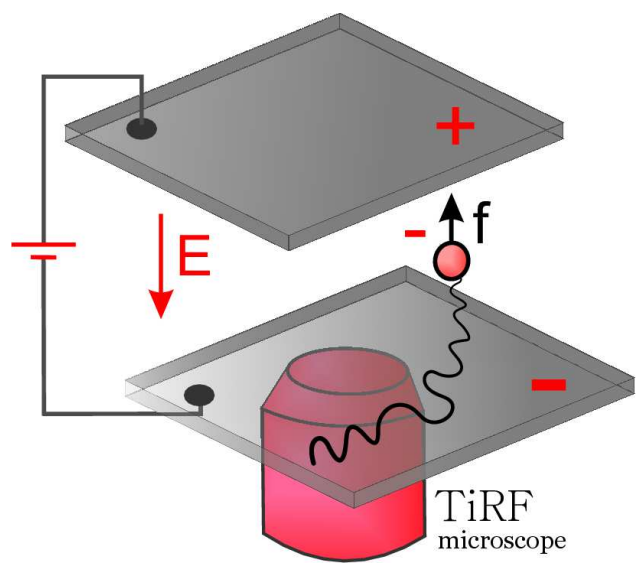}
\caption{Schematic representation of micromanipulation (detachment) experiment
using TIRF microscope.}
\label{fig_tirf}
\end{center}
\end{figure}

One way to do this is by means of {\it total internal reflection fluorescence
microscopy} (TIRFM) which is a powerful technique aimed at imaging fluorescent
species close to an interface ~\cite{Mashanov}. A possible schematic setup is
shown in Fig.~\ref{fig_tirf}

First of all, let us estimate the required tracking forces. In the strong
adsorption limit $a \varphi/k_B T\simeq 0.5$. The Boltzmann constant $k_B = 1.4
\cdot 10^{-23}$~J/K, the temperature $T=300$~K, a bond length $a = 10^{-10}$~m
so
as a result $\varphi = 2.1 \cdot 10^{-11}$~N $=21$~pN. In the case of weak
adsorption, $\varphi \simeq  (k_B T/a N^{\nu})$, that is, for $N = 10^3$ one has
$\varphi \simeq 8.4 \cdot 10^{-13}$~N $\approx 1$~pN. As a result, one deals
typically with forces $1$~pN $\le \varphi \le 20$~pN.

Now, one could estimate which electric field $E$ should be applied to
electrodes, of the capacitor shown in Fig. so as to produce a tracing force $1\;
pN$. The tracking force is $f = q E$, where $q$ is the charge of the ionic
end-group. The typical ion charge $q = 1.6 \cdot 10^{-19}$~C , so that for $E =
5\cdot 10^6$~V/m one gets $f = 5 \cdot  1.6 \cdot 10^{-19} \cdot 10^{6} \approx
10^{-12}$~N $= 1$~pN. This value of the electrical field is much smaller than
the dielectric strength of benzene, $E_{st} = 165 \cdot 10^6$~V/m, or even of
distilled water  $E_{st} = 70 \cdot 10^6$~ V/m \cite{Handbook}, i.e., these
fluids could be used as dielectrics between the plates of the capacitor. We
underline that a totally stretched polymer of chain length, say $N=10^3$, has a
characteristic size of hundred nanometers, so that the distance between plates
of the capacitor should not be larger than  $1 \div 10 \mu m$.
This means that in order to create the electric field $E = 5\cdot 10^6$~V/m one
should apply to the capacitor a voltage $5 \div 50$~V, i.e., no high-voltage
facility
are necessary.

From a chemical point of view, the sample represents a substrate with
end-grafted polymers at a low graft density (i.e., the chains are in a
''mushroom`` conformation). The synthesis of, e.g., polystyrene end-grafted
chains could be performed by {\it surface-initiated radical polymerization}
\cite{Fukuda}, followed by the labeling of the other chain end through carboxyl
groups and fluorescent dyes. These estimations show that an electrostatic
tracking experiment provides a way to test and verify our main theoretical
results.

\section{Conclusion}
\label{sec_summary}

In this work we consider the force-induced desorption of a polymer from a
structureless adhesive surface both analytically and by means of two distinct
methods for computer simulation.

We have shown theoretically that there are three dynamical regimes of a polymer
chain detachment. Depending on the pulling ${\widetilde f}$ and restoring
${\widetilde \varphi}$ forces, one can discriminate between a ''trumpet``
($1/N^{\nu} \ll {\widetilde \varphi} < {\widetilde f} < 1$), ''stem-trumpet``
(${\widetilde \varphi} < 1 < {\widetilde f}$), and ''stem`` ($1 < {\widetilde
\varphi} < {\widetilde f}$) regimes of desorption.

Remarkably, in all these cases the time dependence of the number of desorbed
monomers $M(t)$ and the height of the first monomer (i.e., the  monomer which
experiences the applied external pulling force) $R(t)$ follows an universal
$\sqrt{t}$-law (even though this is {\em not} a diffusion phenomenon). There is,
however, a common physical background with the well-known Lucas-Washburn
$\sqrt{t}$-law of capillary filling~\cite{DDAMLKKB} as with the ejection
kinetics of a polymer chain from a cavity (virus capsid) \cite{AMLKASKB}. In
these seemingly different phenomena there is always a {\em constant} driving
force (meniscus curvature, or polymer entropy) which acts against a gradually
changing drag force (friction) in the course of the process.

We discovered an interesting similarity between the differential equation
governing the monomer density variation $\rho (x, t)$ in time and space, and the
nonlinear porous medium equation (PME) \cite{Vazquez,Barenblatt}. This makes it
possible to use the self-similarity property of PME and derive rigorously the
$\sqrt{t}$-law.

Our extensive MD- and MC-simulations of the detachment kinetics support this
finding as well as the $\langle \tau_d \rangle \propto N^\beta$ scaling with
$\beta = 2$ of mean detachment time $\tau_d$ with chain length $N$. The
theoretically predicted dependence of $\langle \tau_d \rangle$ on pulling force
$f$ and adsorption energy $\epsilon_s$ appears in perfect agreement with the
simulation results. As noted above, the consistency between theory and computer
experiment is well manifested in the case of overdamped dynamics and strong
adsorption. Moreover, by means of MD-simulation we have  shown  that the beads
density distribution plots support the notion of trumpet-like tilted tensile
blob structures. One can envisage not only tilted but, e.g., bended (horn-like)
structures where the local lateral velocity depends on the $x$-coordinate. We
leave this more complicated case for future investigation.

The deviations in the exponent $\beta$ due to inertial effects in the
underdamped dynamical regime ($\beta \approx 1.75$ found in the MD simulation)
as well as close to the adsorption critical point $\epsilon_c$ ($\beta = 1.83$,
in the MC simulation) are also challenging. It appears possible that inertial
effects and fluctuations close to the CAP (both neglected in our theoretical
treatment) may lead to similar consequences for the dynamics of polymer
desorption, manifested by the observed decrease in the detachment exponent
$\beta$. In both cases this suggests that polymer detachment is facilitated.
This is clearly a motivation for further studies of the problem.

\section*{Acknowledgements}

We thank K.L. Sebastian, H.-J. Butt, K. Koynov, and M. Baumgarten for helpful
discussions. A. Milchev is indebted to the Max-Planck Institute for Polymer
Research in Mainz, Germany, for hospitality during his visit and to CECAM-Mainz
for financial support. This work has been supported by the Deutsche
Forschungsgemeinschaft (DFG), grant No. SFB 625/B4.

\newpage
{\bf \Large Polymer Detachment from Adsorbing Surface: Theory, Simulation
and Similarity to Infiltration into Porous Medium }\\
\vskip 0.2 true cm
by Jaroslaw Paturej, Andrey Milchev, Vakhtang G. Rostiashvili, and Thomas A.
Vilgis

TOC graph:
\begin{figure}[ht]
\includegraphics[scale=0.49]{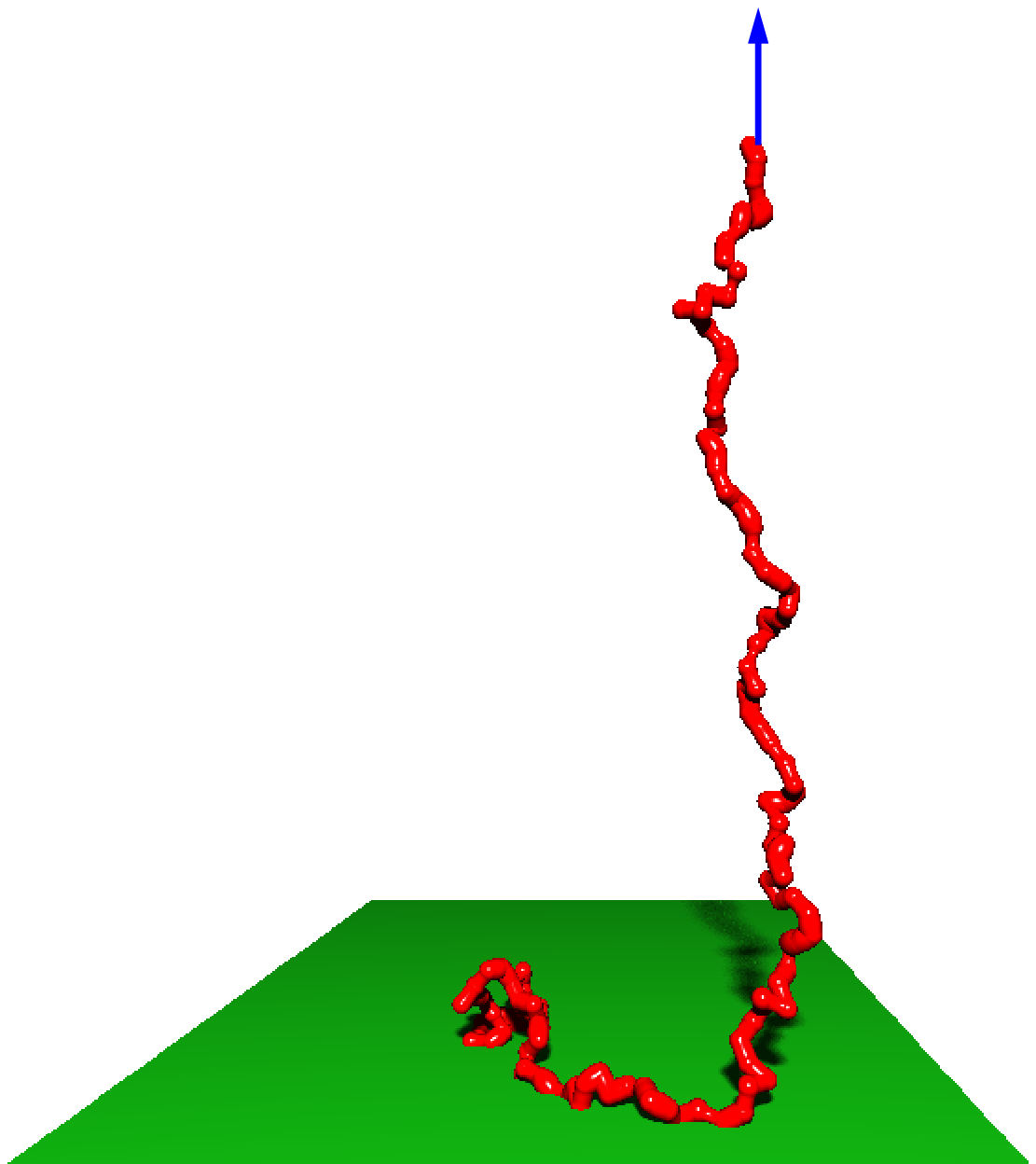}
\caption{A snapshot from MC simulation of a partially desorbed polymer chain
with $N=128$ monomers under pulling force $f = 3.0$ and
surface adhesion $\epsilon_s/k_BT = 2.50$.}\label{TOC_graph}
\end{figure}


\begin{thebibliography}{99}
\bibitem{Ritort}  Ritort, F. {\em J. Phys. Cond. Mat.} {\bf 2006},  {\em 18},
R531.
\bibitem{Somendra}  Bhattacharjee, S. M. {\em J. Phys. A} {\bf 2000}, {\em 33},
L423.
\bibitem{Marenduzzo1} Marenduzzo, D.; Trovato, A.; Maritan, A. {\em Phys. Rev.
E} {\bf 2001}, {\em 64}, 031901.
\bibitem{Orlandini} Orlandini, E.; Bhattacharjee, S.; Mernduzzo, D.; Maritan,
A.; Seno, F. {\em J. Phys. A} {\bf 2001}, {\em 34}, L751.
\bibitem{Kierfeld} Kierfeld, J. {\em Phys. Rev. Lett.} {\bf 2006}, {\em 97},
058302.
\bibitem{Sebastian}  Sebastian, K.L. {\em Phys. Rev. E}  {\bf 2000}, {\em 62},
1128.
\bibitem{Marenduzzo}  Marenduzzo, D.; Bhattacharjee, S.M;  Maritan, A.;
Orlandini, E.;  Seno, F. {\em Phys. Rev. Lett.} {\bf 2002}, {\em 88}, 028102.
\bibitem{SBVRAMTV1} Bhattacharya, S.; Rostiashvili, V. G.; Milchev, A.; Vilgis
T. {\em Phys. Rev. E} {\bf 2009}, {\em 79}, 030802(R).
\bibitem{SBVRAMTV2}  Bhattacharya, S.; Rostiashvili, V. G.; Milchev, A.; Vilgis,
T. {\em Macromolecules} {\bf 2009}, {\em 42}, 2236.
\bibitem{SBAMVRTV}  Bhattacharya, S.;  Milchev, A.; Rostiashvili, V. G.; Vilgis,
T. {\em Eur. Phys. J. E} {\bf 2009}, {\em 29}, 285.
\bibitem{Skvortsov} Skvortsov A. M.; Klushin L. I.; Fleer G. J.; Leermakers,
F. A. M. {\em J. Chem. Phys.} {\bf 2010}, {\em 132}, 064110.
\bibitem{Brochard}  Brochard-Wyart, F. {\em Europhys. Lett.} {\bf 1993}, {\em
23}, 105.
\bibitem{Sakaue_1}  Sakaue,  T. {\em Phys. Rev. E} {\bf 2007}, {\em 76}, 021803.
\bibitem{Sakaue_2}  Sakaue, T. {\em  Phys. Rev. E} {\bf 2010}, {\em 81}, 041808.
\bibitem{Sebastian_1}  Sebastian, K.L.;  Rostiashvili, V.G.;  Vilgis, T.A.;
{\em EPL} {\bf 2011}, {\em 95}, 48006.
\bibitem{Simsen}  Simsen, J.;  Gentile, C.B. {\em Nonlinear Anal.} {\bf 2009},
{\em 71}, 4609.
\bibitem{Vazquez}  V\'azquez, J.L. {\em The Porous Medium Equation}, Clarendon
Press, Oxford, {\bf 2007}.
\bibitem{Barenblatt}  Barenblatt, G.I.;  Entov, V.M.;  Ryzhik, V.M. {\em Theory
of Fluid Flows Through Natural Rocks}, Kluwer Academic Publishers, London, {\bf
1990}.
\bibitem{Gennes}  de Gennes, P.G.; Pincus, P.  {\em J. Phys, (France) Lett.}
{\bf 1983}, {\em 44}, L241.
\bibitem{Gennes_Book}  de Gennes, P.G. {\em Scaling Concept in Polymer Physics},
Cornell
University Press, Ithaca, {\bf 1979}.
\bibitem{Sakaue_3} Sakaue, T.; Yoshinaga, N. {\em Phys. Rev. Lett.} {\bf 2009},
{\em 102}, 148302
\bibitem{Netz} Serr, A.; Netz, R. {\em Europhys. Lett.} {\bf 2006}, {\em 73},
292.
\bibitem{Bhattacharya_1}  Bhattacharya, S.;  Milchev, A.;  Rostiashvili, V.G.;
 Vilgis, T.A. {\em Eur. Phys. J. E}  {\bf 2009}, {\bf 29}, 285.
\bibitem{Bhattacharya_2} Bhattacharya, S.;  Milchev, A.;  Rostiashvili, V.G.;
 Vilgis, T.A. {\em Macromolecules} {\bf 2009},  {\em 42}, 2236.
\bibitem{Vanderzande} Vanderzande, C. {\em Lattice Model of Polymers},
Cambridge University Press, Cambridge, {\bf 1998}.
\bibitem{Lai}  Sheng, Y.-J.;  Lai P.-Y. {\em  Phys. Rev. E} {\bf 1997},  {\em
56}, 1900.
 \bibitem{AMKB}  Milchev, A.; Binder, K. {\em Macromolecules} {\bf 1996}  {\em
29}, 343.
\bibitem{Mashanov} Mashanov G.I.; Tacon D.; Knight A.E.; Peckham M.; Molloy J.
E. {\em Methods}, {\bf 2003}, {\em 29}, 142.
\bibitem{Handbook} {\em CRC Handbook of Chemistry and Physics}, 93rd Edition,
{\bf 2012-2013}.
\bibitem{Fukuda} Y. Tsujii, K. Ohno, S. Yamamoto, A. Goto, T. Fukuda, {\em Adv.
Polym. Sci.} {\bf 2006}, {\em 197}, 1.
\bibitem{DDAMLKKB} Dimitrov, D. I.; Klushin, L.; Milchev, A.; Binder, K. {\em
Phys. Fluids}{\bf 2008}, {\em 20}, 092102.
\bibitem{AMLKASKB} Milchev, A.; Klushin, L.; Skvortsov, A.; Binder K. {\em
Macromolecules} {\bf 2010}, {\em 43}, 6877.

\end{thebibliography}
\end{document}